\documentclass[
reprint,
 amsmath,amssymb,
 aps,
]{revtex4-2}

\usepackage{graphicx}
\usepackage{dcolumn}
\usepackage{bm}
\usepackage{subfigure}

\begin{document}

\preprint{APS/123-QED}

\title{On point-ahead angle control strategies for TianQin}

\author{Dezhi Wang}
\author{Xuefeng Zhang}
\email{zhangxf38@sysu.edu.cn}
\author{Hui-Zong Duan}
\email{duanhz3@sysu.edu.cn}

\affiliation{MOE Key Laboratory of TianQin Mission, TianQin Research Center for Gravitational Physics $\&$ School of Physics and Astronomy, Frontiers Science Center for TianQin, Gravitational Wave Research Center of CNSA, Sun Yat-sen University (Zhuhai Campus), Zhuhai 519082, China}

\date{\today}

\begin{abstract}

Pointing-related displacement noises are crucial in space-based gravitational wave detectors, where point-ahead angle control of transmitted laser beams may contribute significantly. For TianQin that features a geocentric concept, the circular high orbit design with a nearly fixed constellation plane gives rise to small variations of the point-ahead angles within $\pm 25$ nrad in-plane and $\pm 10$ nrad off-plane, in addition to a static bias of 23 $\mu$rad predominantly within the constellation plane. Accordingly, TianQin may adopt fixed-value compensation for the point-ahead angles and absorb the small and slow variations into the pointing biases. To assess the in-principle feasibility, the far-field tilt-to-length (TTL) coupling effect is discussed, and preliminary requirements on far-field wavefront quality are derived, which have taken into account of TTL noise subtraction capability in post processing. The proposed strategy has benefits in simplifying the interferometry design, payload operation, and TTL noise mitigation for TianQin. 

\end{abstract}

\maketitle

\section{Introduction} \label{sec:intro}

\begin{figure}[!h]
\includegraphics[width=0.45\textwidth]{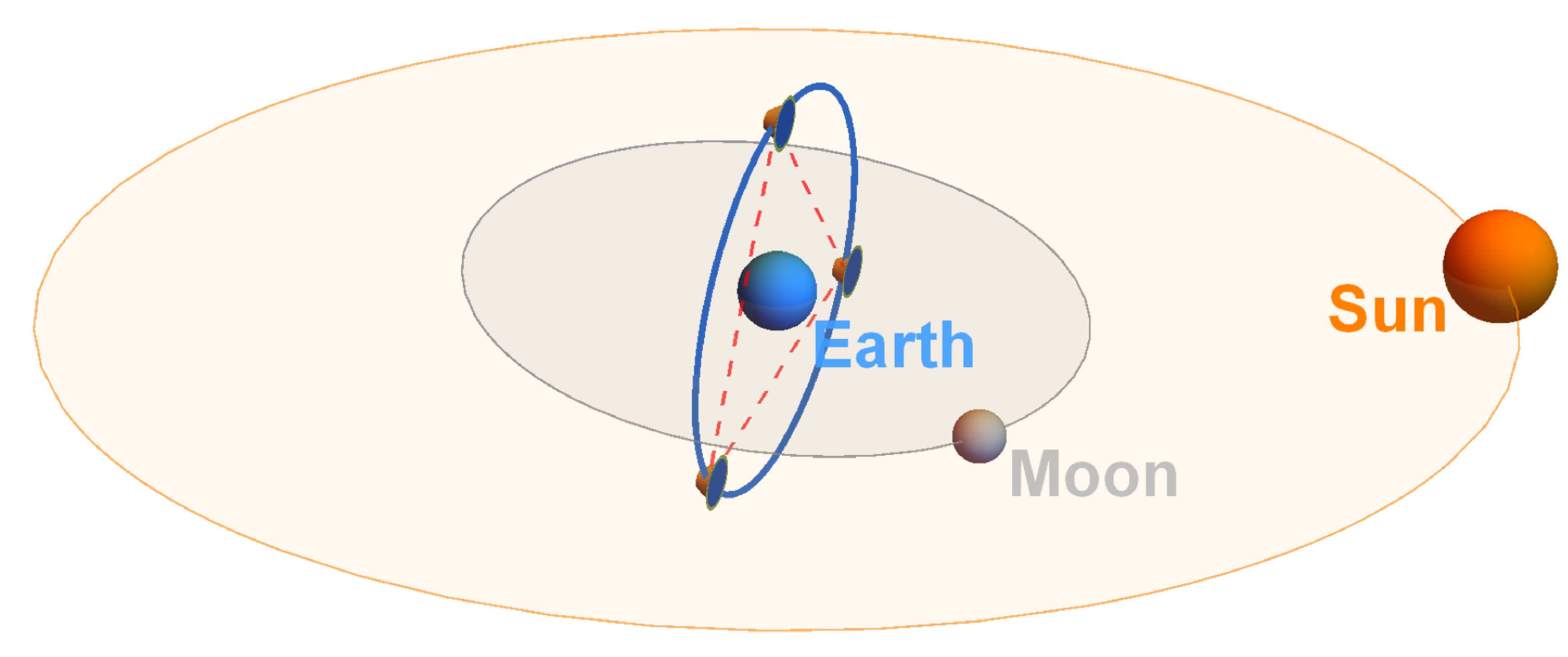}
\caption{\label{fig:TQ_orbit} A schematic diagram of TianQin's orbit and constellation in the Earth-centered ecliptic reference frame. }
\end{figure}

TianQin is a proposed space-based gravitational-wave (GW) detection mission to operate in high Earth orbits \cite{Luo2016}. The detector consists of three identical drag-free controlled satellites to form a nominal equilateral triangle with an armlength of $\sim 1.7 \times 10^{5}$ km (see Fig. \ref{fig:TQ_orbit}). The mission design also features a nearly fixed constellation plane (hence the detector pointing also fixed) that is set almost perpendicular to the ecliptic plane. Heterodyne laser interferometric links are to be established between free-falling test masses (TMs) inside different satellites to monitor picometer-level armlength changes in the frequency band of 0.1 mHz--1 Hz. 

A forerunner in space-based GW detection is the LISA mission \cite{LISA2017}, which uses heliocentric orbits trailing or leading the Earth by $\sim 20^\circ$, and has an armlength of $2.5\times 10^6$ km. By a special arrangement of the orbits, the plane of the triangular constellation can revolve around the Sun and form a $60^\circ$ angle relative to the sunlight, providing a stable thermal environment for the three spacecraft. Moreover, the variable detector pointing may also help improve the sky localization of GW sources.

The payload design of LISA has put forward that every spacecraft carries two Movable Optical Sub-Assemblies (MOSAs), each consisting of a telescope, an optical bench, and an inertial sensor, rigidly connected to one another by supporting structures \cite{weise2017optical}. The TM inside the inertial sensor is kept free-floating along the interferometry arm and suspended in other degrees of freedom by electrostatic forces. The entire MOSAs can rotate in the constellation plane via tracking mechanisms to correct for the internal angle variation ($\pm 1^\circ$) of the triangle due to orbital dynamics. Meanwhile, the Drag-free and Attitude Control (DFAC) of the spacecraft is responsible for tracking the pointing variation orthogonal to the constellation plane with micro-Newton thrusters \cite{LISA2000}. The applicability of such a payload architecture and control strategy to TianQin has been studied, and an adaptation to the geocentric orbits can be made \cite{Fang2023}. 

\begin{figure}[htbp]
\includegraphics[width=0.45\textwidth]{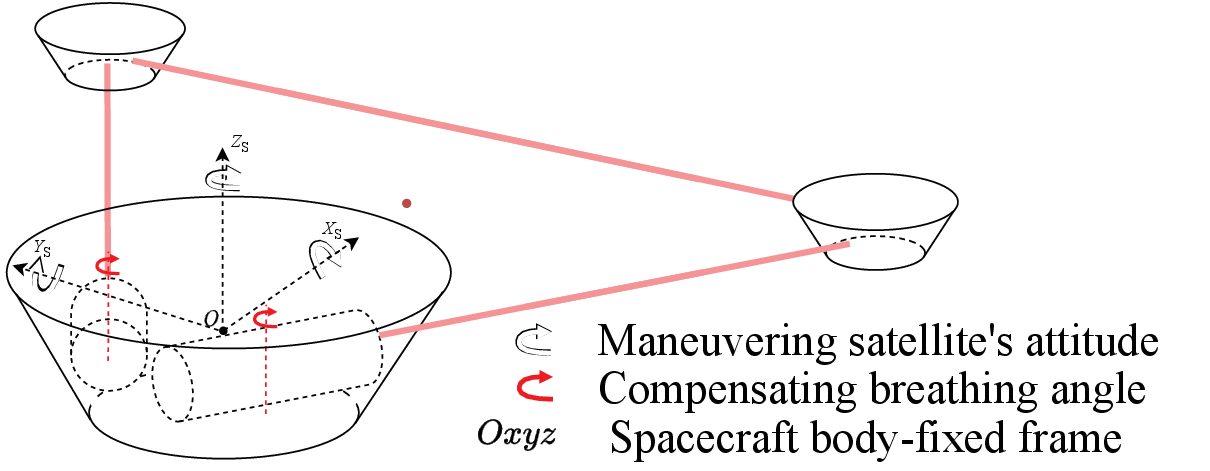}
\caption{\label{fig:TQ_pointing} The overall pointing scheme for TianQin. The off-plane degrees of freedom are tracked by DFAC, and the in-plane angle variation by single-axial MOSA rotation. }
\end{figure}

Accurate pointing of outgoing laser beams at distant spacecraft is crucial for both LISA and TianQin since pointing errors can have a major impact on the precision of the TM-to-TM displacement measurement. Estimated pointing requirements are $\sim 10$ nrad in DC bias and $\sim 10$ nrad/Hz$^{1/2}$ in jitters \cite{LISA2017}, which push the boundary of the current state of the art. Given the MOSA configuration and the involvement of DFAC, the issue of fine-pointing should also be dealt with on the system level.

Due to finite light travel time, a complication in pointing control arises where a certain amount of leading ahead is needed for the transmitted beams to compensate for the lateral motion of the remote satellite relative to the line of sight. The point-ahead angle (PAA) refers to the angle between the received and transmitted beams \cite{LISA98}, which can be decomposed into in-plane and off-plane components when projecting onto the constellation plane. The values of PAA can be fully determined by the orbits and must be taken into account in formulating pointing control strategies to meet the stringent requirements.

The constellation plane of LISA features yearly variation around the Sun, and when combined with the cartwheel motion of the triangle, it gives rise to an off-plane PAA variation of $\pm 3.4\ \mu$rad about the nearly zero mean over the course of one year, as well as an in-plane bias of $\sim 1.7\ \mu$rad with $\pm 10$ nrad variation \cite{Houba2022}. The comparatively large off-plane variation necessitates a dedicated PAA mechanism on the optical bench for in-orbit compensation in either open-loop or closed-loop manners \cite{LISA2011}. Since the mechanism is located in the optical path of the outgoing laser beam, it has to meet strict requirements on path-length and angular stabilities \cite{henein2009design}, which may also be monitored by a dedicated interferometer on the optical bench \cite{Jennrich2009CQG}.

As hinted earlier, an important issue faced by PAA control is the tilt-to-length (TTL) coupling, which refers to the effect that jitters of the satellites and MOSAs can induce displacement noise in the longitudinal interferometric measurements between two TMs. For small jitters near a given angle, the effect can be modeled to the first order by a linear relationship via a set of coupling coefficients (TTL factors) \cite{Houba22JGCD}. Due to inevitable misalignment and imperfection of optical components and pointing offsets of laser beams, it is highly challenging to reduce the coupling coefficients to desired small values in hardware, and therefore dedicated TTL factor estimation and noise subtraction in post-processing seem necessary in the current measurement scheme \cite{Houba22JGCD,Paczkowski22PRD,Houba23CQG,George2023}. Unfortunately for various reasons, PAA adjustments are expected to change the TTL factors \cite{Houba2022,Hasselmann2021}. The temporal variability complicates the estimation processes of the TTL factors, which generally require sufficiently long windows of data streams to reach the desired accuracy \cite{Houba2022,George2023}. To tackle the issue, an open-loop control strategy for discrete PAA corrections has been proposed for LISA \cite{Houba2022} which maximizes intervals between adjustments through coordination of the six optical benches. Nevertheless, more studies appear to be needed to demonstrate realistically the impact of PAA adjustments on the system performance and the quality of data products.

Regarding TianQin, the situation is different, as we will show that the PAA variation is rather small owing to its geocentric co-planar orbit design with a nearly fixed constellation plane in the inertial space (Sec. \ref{sec:orbits_PAA}). Therefore, it motivates us to pursue a different PAA control strategy, where only the static bias is compensated. To examine its feasibility, we analyze the far-field TTL effect due to the uncorrected PAA variation and discuss possible countermeasures (Sec. \ref{sec:strategyTTL}). In addition, a few prominent aspects of implementation are commented (Sec. \ref{sec:conclusion}). We expect that the strategy can help circumvent potential difficulties in TTL mitigation with time-variable coupling coefficients due to PAA adjustments.

\section{TianQin's orbits and point-ahead angles} \label{sec:orbits_PAA}

\begin{figure}[!ht] 
    \centering
    \includegraphics[width=0.45\textwidth]{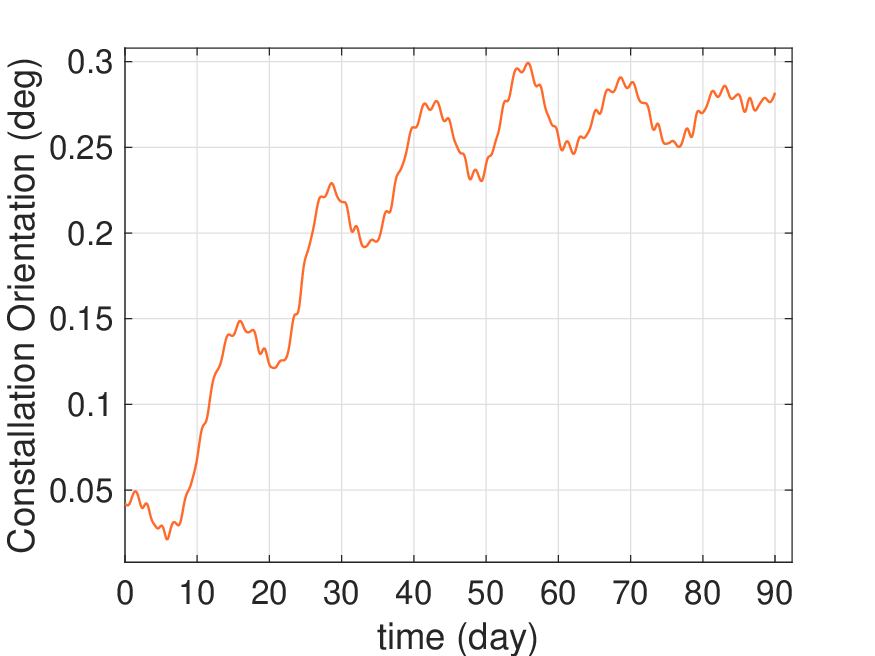}
    \caption{\label{fig:plane_drift} The angle variation between the normal of the TianQin constellation and the nominal direction to the reference GW source, i.e., the white dwarf binary RX J0806.3+1527 \cite{Ye2019}.}
\end{figure}

One can calculate the PAA evolution from the orbits. A set of optimized initial orbital elements is given in Table \ref{tab:orbit} \cite{Ye2019}, taking account of lunisolar perturbation and the non-spherical gravity of the Earth. Assuming pure gravity flight under drag-free control, the ensuing orbital evolution fulfills the constellation stability requirements, e.g., the breathing angle variation within $60\pm 0.1^\circ$. Owing to the high altitude, the constellation plane is nearly fixed with only a small variation of $\sim 0.05^\circ$ per orbit (3.6 days, see Fig. \ref{fig:plane_drift}). 

\begin{table}[h]
\caption{\label{tab:orbit} Optimized initial orbital elements of TianQin satellites (SC1, 2, 3) in the J2000-based Earth-centered ecliptic reference frame, including the semimajor axis $a$, the eccentricity $e$, the inclination $i$, the longitude of the ascending node $\Omega$, the argument of periapsis $\omega$ and the true anomaly $\nu$ at 22 May 2034 12:00:00 UTC \cite{Ye2019}.}
\begin{ruledtabular}
\begin{tabular}{cccc}
     \ & SC1& SC2& SC3 \\
     \hline
     $a$ (km)& 99995.572323&100011.400095& 99993.041899 \\
     $e$&0.000430&0.000000&0.000306\\
     $i(^{\circ})$&94.697997&94.704363&94.709747\\
     $\Omega(^{\circ})$&210.445892&210.440199&210.444582\\
     $\omega(^{\circ})$&358.624463&0.000000&0.001624\\
     $\nu(^{\circ})$&61.329603&179.930706&299.912164\\
\end{tabular}
\end{ruledtabular}
\end{table}

\begin{figure}[!ht]
    \centering
    \includegraphics[width=0.45\textwidth]{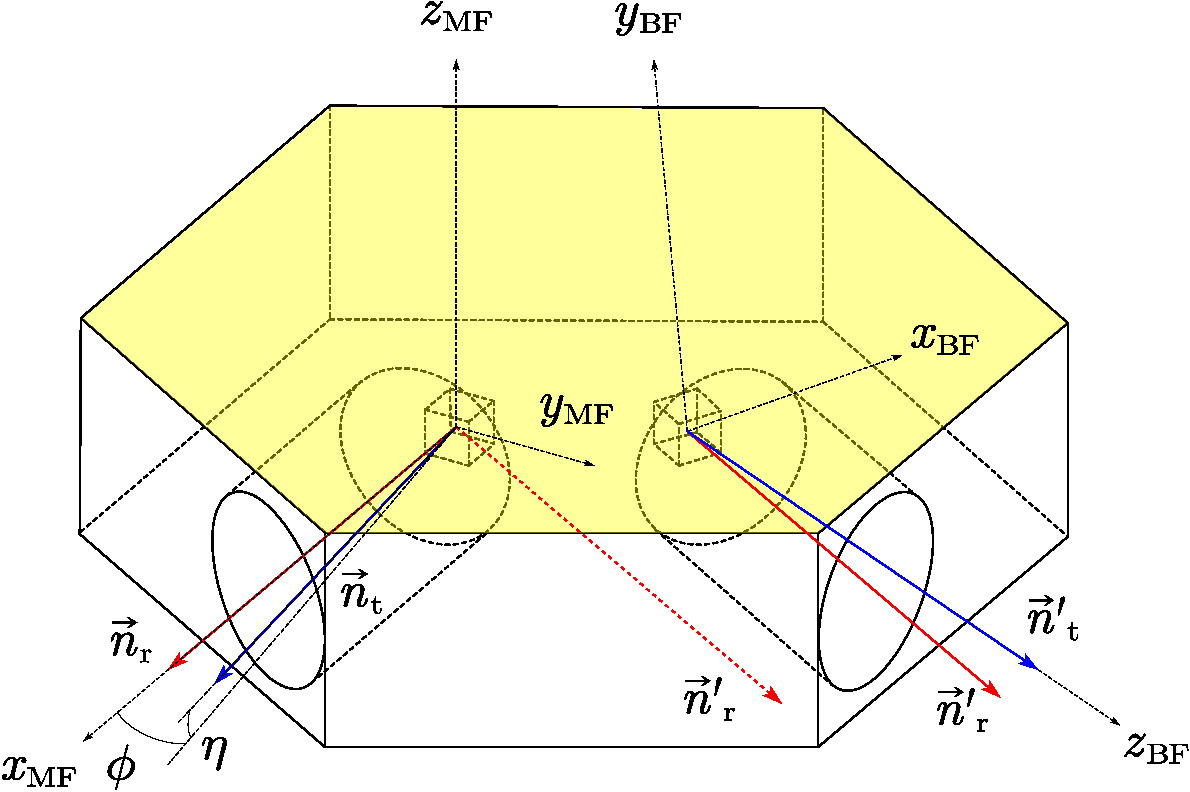}
    \caption{\label{fig:SF} A schematic diagram of the MOSA frame (left) and the beam frame (right, see Sec. \ref{sec:strategyTTL}). The vector $\vec{n}_{\mathrm{r}}$ denotes the opposite direction of the incoming beam, and $\vec{n}_{\mathrm{t}}$ denotes the transmitted direction. The primed notations refer to the corresponding vectors of the adjacent MOSA. The angles $\phi$ and $\eta$ denote the in-plane PAA and off-plane PAA, respectively. }
\end{figure}

Directions of transmitted beams and PAAs can be described in the MOSA frames (MF), as illustrated in Fig. \ref{fig:SF}. The origin of a MOSA frame is at the geometric center of the electrode housing of the TM, which coincides with the TM's center of mass (CoM) in the nominal state. The MOSA axis ($x_\mathrm{MF}$) is aligned with the received beam, which is realized by nulling the differential wavefront sensing (DWS) signal. The DWS measures the pitch and yaw angles between the wavefronts of the local reference beam and the received beam using quadruple photodiodes, and provides the input for the pointing/attitude control. The two MOSA axes aligned with the two received beams define what we call the local constellation plane. Then the $y_\mathrm{MF}$-axis is defined to be in the plane (see Fig. \ref{fig:SF}) and point to the adjacent MOSA, so that the direction of the transmitted beam can be uniquely specified with respect to the received beam. Thereby PAAs can be decomposed into in-plane and off-plane components. In calculating PAAs, we neglect the separation of the TM's CoM from the satellite's CoM ($\sim 0.2$ m), given the huge armlength of TianQin. 

\begin{figure}[!ht]
\includegraphics[width=0.45\textwidth]{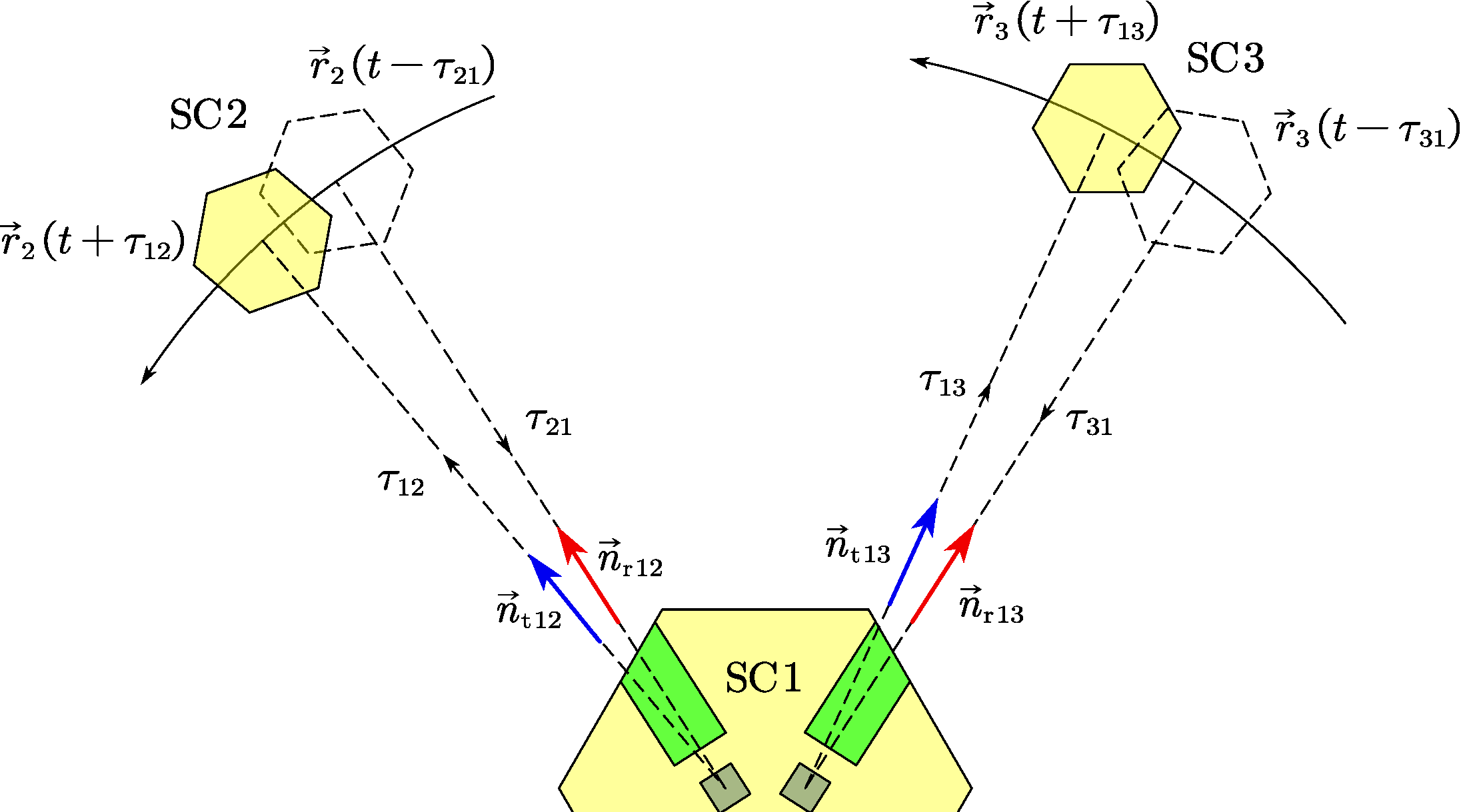}
\caption{\label{fig:TQ_PAA} A sketch of TianQin's PAAs. }
\end{figure}

\begin{figure*}[!htb]
    \centering
    \subfigure[]{\label{fig:phi12}\includegraphics[width=0.45\textwidth]{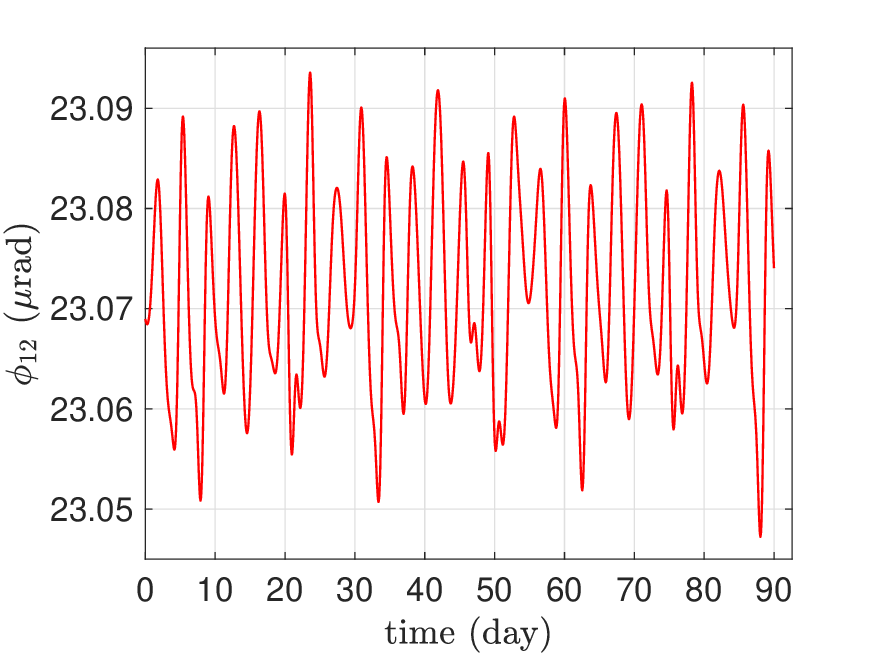}}
    \subfigure[]{\label{fig:phi13}\includegraphics[width=0.45\textwidth]{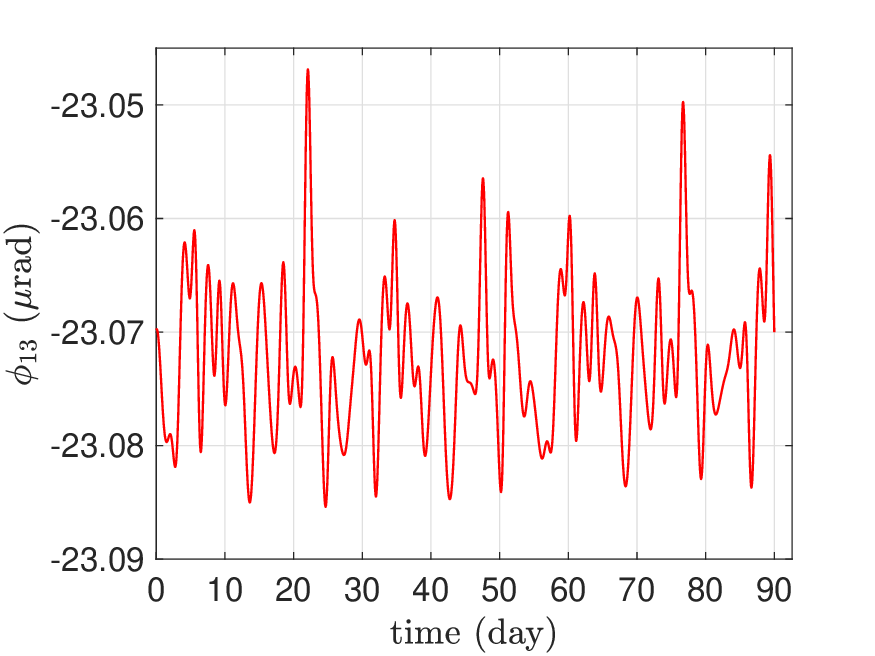}}
    \subfigure[]{\label{fig:eta12}\includegraphics[width=0.45\textwidth]{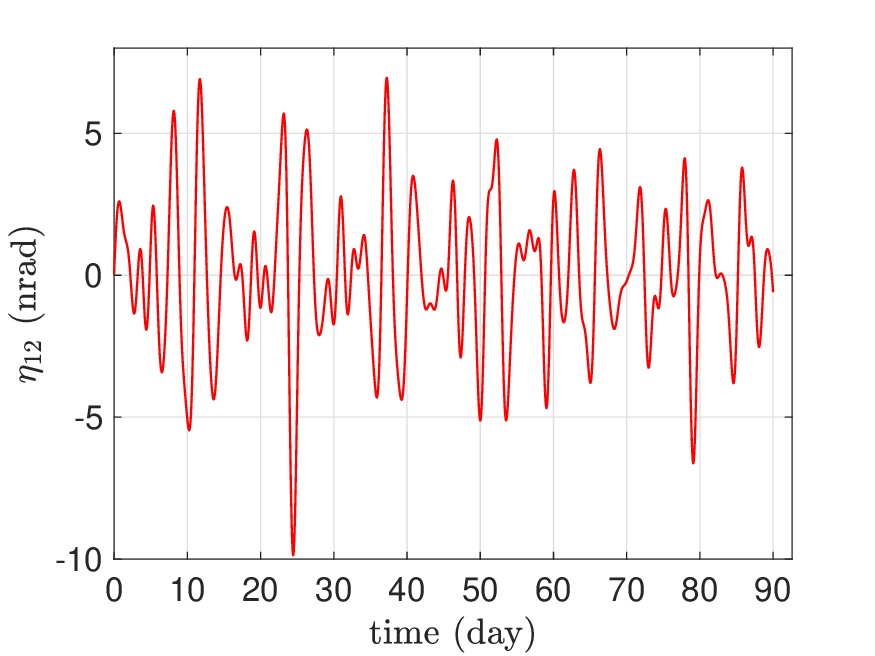}}
    \subfigure[]{\label{fig:eta13}\includegraphics[width=0.45\textwidth]{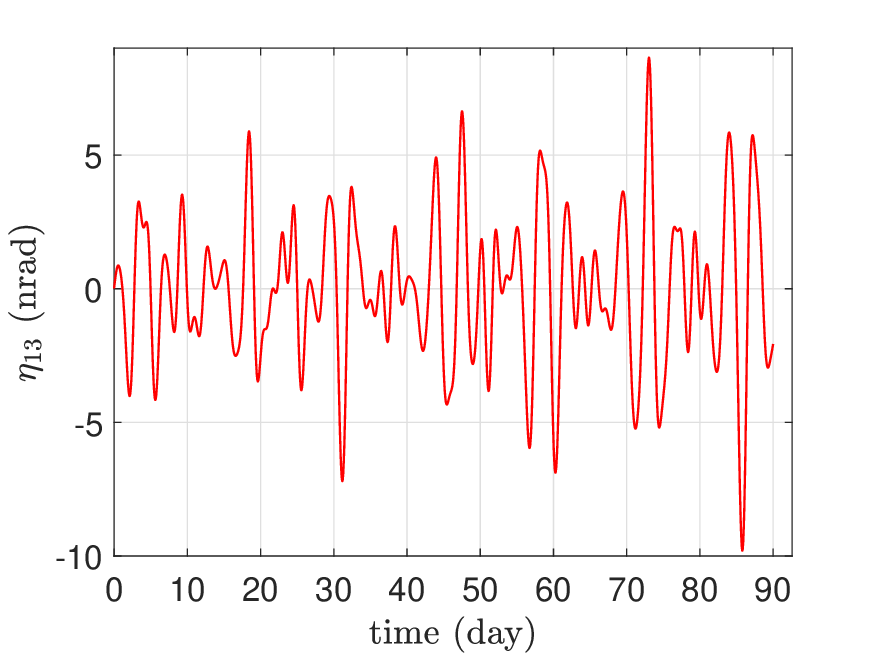}}
    \caption{ The plots (a) and (b) illustrate the evolution of the in-plane components of PAA$_{12}$ and PAA$_{13}$, respectively. Likewise for the plots (c) and (d) to show the off-plane components. }
    \label{fig:PAA}
\end{figure*}

The situation of PAAs is further illustrated in Fig. \ref{fig:TQ_PAA}, where the vector $\vec{n}_{\mathrm{t}ij}$ represents the direction of the beam from SC$i$ to SC$j$ ($i,j=1,2,3$), and $\vec{n}_{\mathrm{r}ij}$ represents the opposite direction of the beam from SC$j$ to SC$i$. Therefore the PAA can be defined as the angle between $\vec{n}_{\mathrm{r}ij}$ and $\vec{n}_{\mathrm{t}ij}$. The propagation time $\tau_{ij}$ for the beam traveling from SC$i$ to SC$j$ can be determined from
\begin{equation}
    \tau_{ij}=\vert \vec{r}_{j}(t+\tau_{ij}) - \vec{r}_{i}(t) \vert /c,
\end{equation} 
and similarly for $\tau_{ji}$ from
\begin{equation}
    \tau_{ji}=\vert \vec{r}_{j}(t-\tau_{ji}) - \vec{r}_{i}(t) \vert /c,
\end{equation} 
where $c$ is the light speed and we have neglected the general relativistic effect. Solving these implicit equations iteratively yields $\tau_{ij}$ and $\tau_{ji}$, and hence $\vec{n}_{\mathrm{t}ij}$ and $\vec{n}_{\mathrm{r}ij}$ can be calculated. We use PAA$_{ij}$ to denote the PAA of SC$i$ with respect to SC$j$ and use $\phi$ and $\eta$ to denote the in-plane and off-plane PAAs, respectively. Utilizing simulated TianQin orbit data, the PAA evolutions of SC1 during a 3-month observation period are computed and shown in Fig. \ref{fig:PAA}. 

\begin{figure}[htbp]
    \centering
    \includegraphics[width=0.45\textwidth]{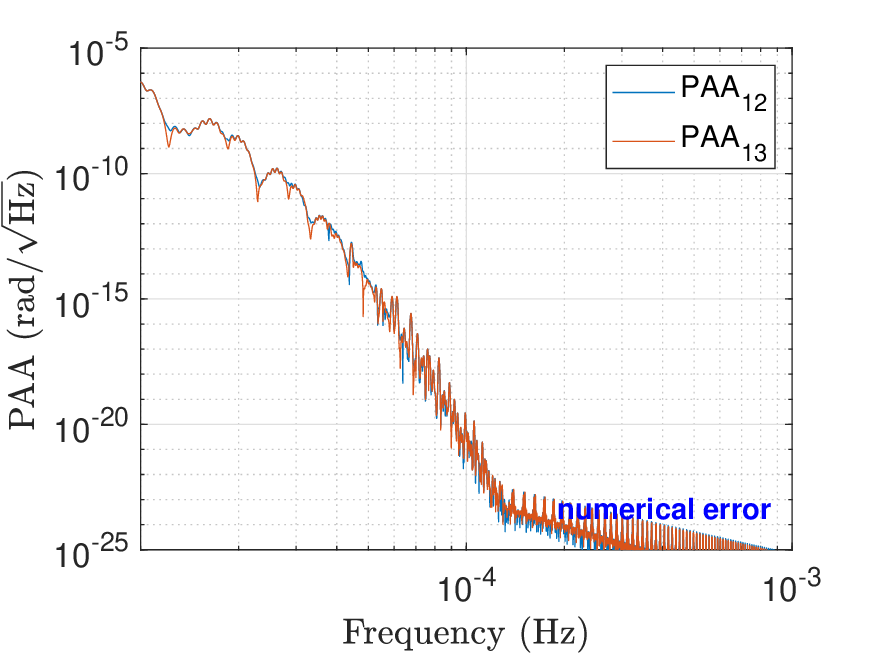}
    \caption{\label{fig:PAAASD} Amplitude spectral densities (ASD) of PAA$_{12}$ and PAA$_{13}$. The flattened parts of the ASD curves above $1.3 \times 10^{-4}$ Hz are due to numerical error. } 
\end{figure}

In Fig. \ref{fig:PAA}, the off-plane PAAs of TianQin are characterized by fluctuations centering at 0 with a range of $\pm 10$ nrad, while the in-plane components oscillate around +23.07 $\mu$rad or $-23.07$ $\mu$rad with a range of $\pm 25$ nrad. The PAA variations are quite small compared to the divergence angle of the outgoing beam ($\sim 5 \mu$rad), which suggests that the remote satellite can still receive sufficient laser intensity even with a fixed PAA. Furthermore, Fig. \ref{fig:PAAASD} reveals that the PAA is slowly varying, and has negligible contribution to pointing jitters in the detection band.

\section{Proposed strategy and far-field TTL effect} \label{sec:strategyTTL}

Regarding TianQin's PAA control, since the variation is quite small, we envision a fixed offset compensation without active real-time adjustment at least in the science mode. Therefore, one needs to absorb the small and slow PAA variations ($\pm 25$ nrad) into the pointing biases (10 nrad DC) of the laser beams. Given increased total pointing biases (up to $\sim 35$ nrad), the requirement on the far-field wavefront quality becomes more stringent in order to keep the coupling noise of the wavefront error and pointing jitters at roughly the same level \cite{Bender2005}. In this section, we analyze this far-field TTL effect and estimate corresponding requirements.

\subsection{Gaussian beam model}

Implementing the static compensation strategy of PAAs makes the optical field of the transmitted beam stationary relative to the transmitting MOSA. However, it introduces extra motion of the receiving satellite relative to the distant optical field. In other words, the remote satellite may swing back and forth in the field of view of the emitting satellite. This relative motion due to uncompensated PAA may induce an additional TTL effect. To analyze the impact, a simulation is conducted to track the positions of SC2 and SC3 with respect to the transmitted beams from SC1 during one 3-month observation period.

To describe the motion of the remote satellite relative to the transmitted beam, we introduce the beam frame (BF). As depicted in Fig. \ref{fig:SF}, the $z_{\mathrm{BF}}$-axis is the transmitted beam direction with a fixed PAA relative to the incoming beam direction, and the $x_{\mathrm{BF}}$-axis is aligned with the plane containing both received and transmitted directional vectors ($\vec{n}'_\mathrm{r}$, $\vec{n}'_\mathrm{t}$), and the $y_{\mathrm{BF}}$-axis follows the right-hand rule.

For an initial setup, the laser field between two satellites is modeled by an ideal Gaussian beam propagating along the $z_{\mathrm{BF}}$-axis. Due to the rotational symmetry about $z_{\mathrm{BF}}$, we use the cylindrical coordinates $\left( \rho, \varphi, z \right)$ instead. The electric component of the Gaussian beam with the waist at the origin is given by
\begin{equation}
\label{eq:Gaussian}
    \vec{E} = \vec{E}_{0} \sqrt{\frac{2}{\pi}} \frac{1}{w(z)} \mathrm{e}^{- \frac{\rho^{2}}{w(z)^{2}}} \mathrm{e}^{\mathrm{i} \left[ \omega t - \left( kz - \phi_G (z) + \frac{k\rho^{2}}{2R(z)} \right) \right]}
\end{equation}
where $w(z) = w_0 \sqrt{1 + (z/z_R)^2}$ gives the beam radius, and $\phi_G (z) = \arctan \left( z/z_R \right)$ denotes the Gouy phase shift, and the wavefront curvature reads $R(z) = z + z_{R}^2/z$. Moreover, $k=2\pi/\lambda$ is the wave number with the wavelength $\lambda = 1064$ nm, and $\omega$ is the angular frequency. The quantity $z_R=\pi w_0^2/\lambda$ is the Rayleigh range \cite{Freise2010}, and for TianQin, we have the waist radius $w_0=13.3$ cm. The flight time taken by the wavefront emitted from the waist to reach the point $\left( \rho, \theta, z \right)$ is denoted by $\tau \left( \rho, z \right)$, which can be derived from the phase term of Eq. (\ref{eq:Gaussian}) as
\begin{equation}
    \tau \left( \rho, z \right) = \frac{kz - \phi_G (z) + \dfrac{k\rho^2}{2R(z)}}{\omega}.
\end{equation}

\begin{figure}[htbp]
    \centering
    \includegraphics[width=0.45\textwidth]{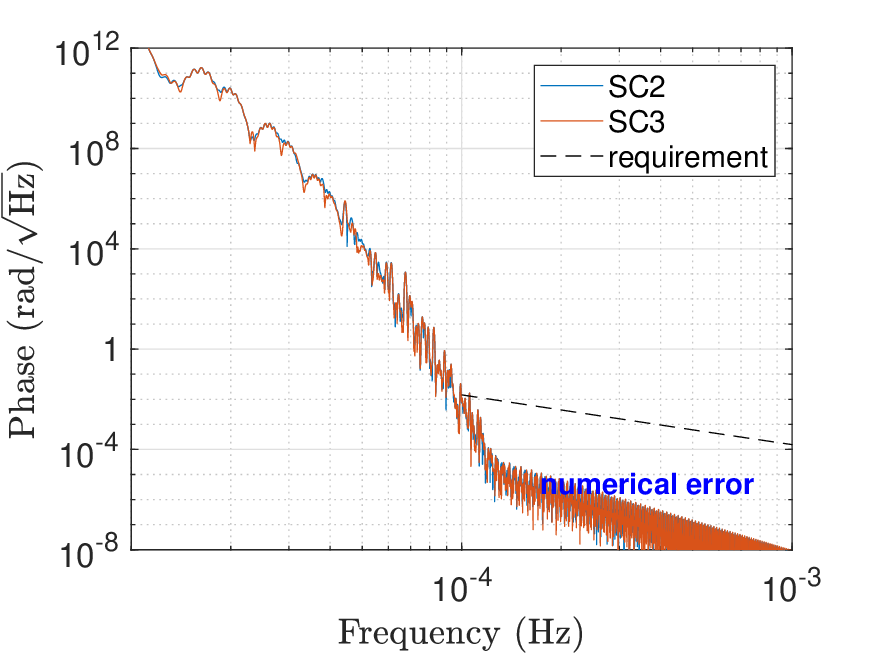}
    \caption{ASD curves of the phases received by SC2 and SC3 from SC1 compared with the requirement of TianQin. The low-frequency bulge is due to orbital perturbation \cite{Zhang2021}. The flattened parts of the ASD curves above $1.3\times 10^{-4}$ Hz are due to numerical error.}
    \label{fig:WFASD}
\end{figure}

The wavefront is emitted at $t_1$ and reaches the remote satellite at $t_2$. The position of the remote satellite at $t_2$ in the beam frame at $t_1$ can be expressed as $\left( \rho(t_2, t_1), \theta(t_2, t_1), z(t_2, t_1) \right)$. The relation between $t_1$ and $t_2$ is determined by
\begin{equation} \label{eq:propagate}
    t_2-t_1=\tau\left( \rho\left(t_2, t_1\right), z\left(t_2, t_1\right) \right)
\end{equation}
which allows us to solve for $t_1$ given each $t_2$ by iteration, and vice versa. Subsequently, the received phase at the remote satellite can be computed. The Fig. \ref{fig:WFASD} shows that the received phase fluctuation due to the static PAA compensation predominantly distributes below $1\times 10^{-4}$ Hz due to the orbital dynamics, and that the impact is considerably lower than TianQin's requirement in the case of Gaussian beams. The ASD result is consistent with the estimated Doppler effect due to luni-solar gravitational perturbation \cite{Zhang2021,Zheng2023}. 

\begin{figure}
    \centering
    \includegraphics[width=0.45\textwidth]{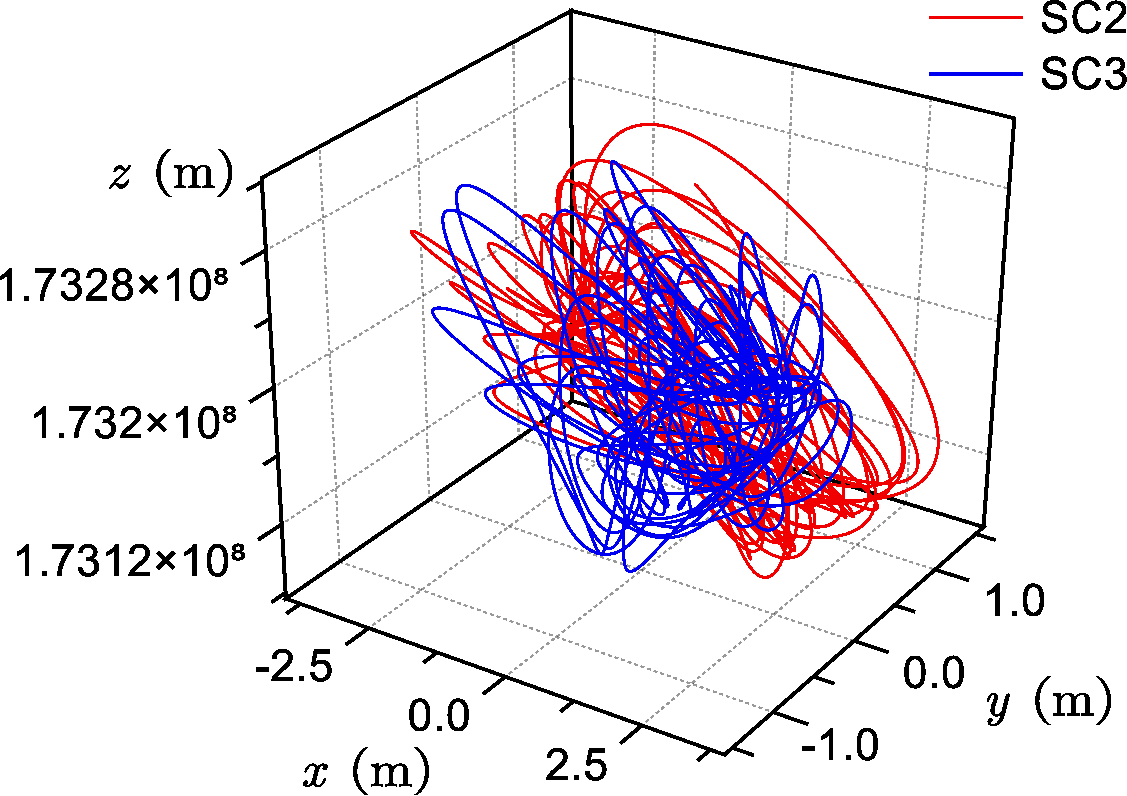}
    \caption{The traces of SC2 and SC3 in the beam frames of SC1 for 3 months. Here $z$ is the axis of the emitted Gaussian beam. }
    \label{fig:trace}
\end{figure}

Additionally, we can obtain the trace of the remote satellite in the beam frame (see Fig. \ref{fig:trace}), and the lateral satellite motion due to uncompensated PAAs is only of a few meters in the $x$-$y$ plane. Based on this initial model, we can further substitute the ideal Gaussian beam with distorted Gaussian beams in the beam frame and analyze the resulting far-field TTL effect. 

\subsection{Effect of far-field wavefront error}

\begin{figure}[h]
    \centering
    \includegraphics[width=0.45\textwidth]{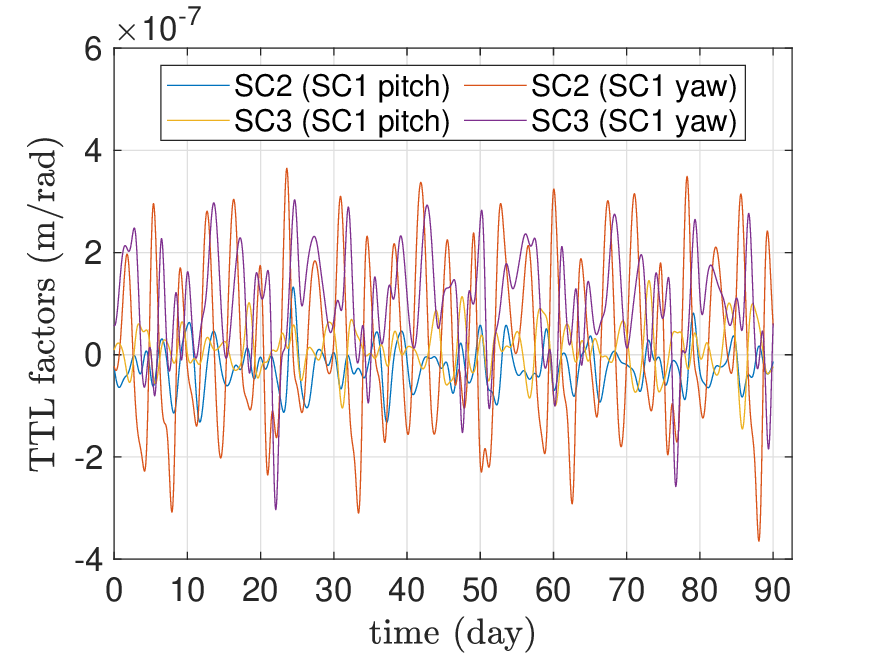}
    \caption{The TTL factors induced by ideal Gaussian beams received at SC2 and SC3 during one 3-month observation window. }
    \label{fig:TTLfac}
\end{figure}

\begin{figure}[hbp]
    \centering
    \includegraphics[width=0.45\textwidth]{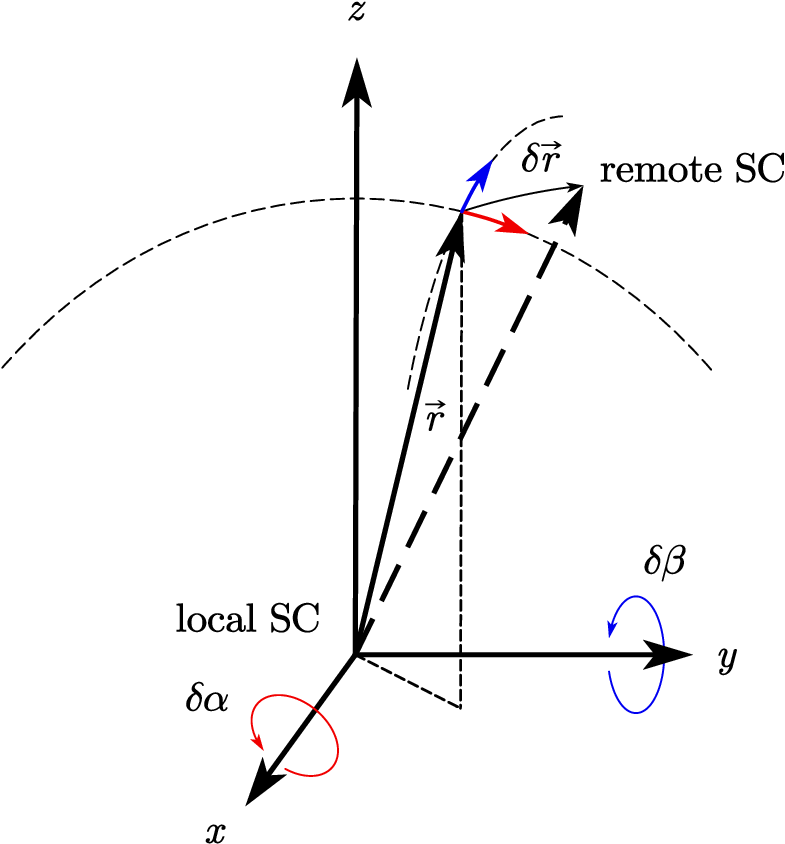}
    \caption{\label{fig:jitterwithBF} Owing to the local pointing jitters ($\delta\alpha$, $\delta\beta$), the remote satellite has a relative displacement $\delta \vec{r}$ in the beam frame $(x,y,z)$. The far-field wavefront distortion along $\delta \vec{r}$ can couple into the phase read-out.}
\end{figure}

Previous studies have revealed that the wavefront of an emitted beam is not ideally spherical but distorted in reality, which leads to additional phase noise at the receiving satellite when combined with pointing jitters of the emitting beam (see, e.g., \cite{Bender2005,Sasso2018,Ming2021,Weaver2022,Xiao2023}). Various factors including aberration in the optical system contribute to this phase deviation which is denoted by $w_{e}(x,y,z)$. Thereby we model the far-field phase by a scalar field in Cartesian coordinates, i.e. 
\begin{equation} \label{eq:phase}
    \Phi(x,y,z)=kz - \phi_G (z) + \frac{k(x^2+y^2)}{2R(z)} +w_{e}(x,y,z), 
\end{equation} 
where we have dropped the time-dependent term $\omega t$. The displacement noise induced by pointing jitters can be approximated by
\begin{equation} \label{eq:measuringnoise}
    \delta L = \frac{1}{k} \nabla \Phi \cdot \delta \vec{r}, 
\end{equation}
where $\nabla \Phi$ is the phase gradient and $\delta \vec{r}$ is the relative displacement of the remote satellite in the beam frame due to jitters (see Fig. \ref{fig:jitterwithBF}).  Given the remote satellite position $\vec{r} = (x,y,z)$, the displacement $\delta \vec{r}$ can be expressed as
\begin{equation} \label{eq:displacement}
    \delta \vec{r} = -z \delta \beta \vec{e_x}+ z \delta \alpha \vec{e_y}+(x \delta\beta-y \delta\alpha)\vec{e_z}, 
\end{equation}
where $\delta \alpha$ and $\delta \beta$ represent the angular jitters around the $x$-axis (pitch) and the $y$-axis (yaw), respectively. The wavefront error $w_e$ is expected to have a rather weak dependence on $z$ at the far field, and hence we neglect the term $\partial w_e/ \partial z$. Substituting Eq. (\ref{eq:displacement}) into Eq. (\ref{eq:measuringnoise}), the TTL factors along two orthogonal directions can be obtained as
\begin{widetext}
    \begin{equation}
     \frac{\partial L}{\partial \alpha} = \frac{z}{k}\left(\frac{\partial}{\partial y}w_{e}\! \left(x,y,z\right)\right)+\frac{\left(-2 k z_{R}^{4}+2 z_{R}^{3}-k \left(x^{2}+y^{2}+2 z^{2}\right) z_{R}^{2}+2 z_{R} z^{2}+k \,z^{2} \left(x^{2}+y^{2}\right)\right) y}{2 k \left(z^{2}+z_{R}^{2}\right)^{2}}
     \label{eq:TTLalpha}
\end{equation}
and
\begin{equation}
    \frac{\partial L}{\partial \beta} =- \frac{z}{k} \left(\frac{\partial}{\partial x}w_{e}\! \left(x,y,z\right)\right)-\frac{x \left(-2 k z_{R}^{4}+2 z_{R}^{3}-k \left(x^{2}+y^{2}+2 z^{2}\right) z_{R}^{2}+2 z_{R} z^{2}+k \,z^{2} \left(x^{2}+y^{2}\right)\right)}{2 k \left(z^{2}+z_{R}^{2}\right)^{2}}.
    \label{eq:TTLbeta}
\end{equation}
\end{widetext}
On the right hand sides of the above equations, the first terms represent the TTL factors due to the wavefront distortion $w_e$, and the second terms are due to the Gaussian wavefront. Because the Gaussian contribution (see Fig. \ref{fig:TTLfac}) is significantly lower than the requirement of $2\times 10^{-5}$ m/rad, we can neglect the second terms in the subsequent analysis. 

In the current measurement scheme, the TTL effect can be calibrated and subtracted from the science data in post-processing (see, e.g., \cite{Paczkowski22PRD}). This can help alleviate the requirements on the far-field wavefront quality. Assuming a conservative capability of subtracting the TTL noise to a residual level of $20\%$ ($\sim 5\%$ in \cite{Paczkowski22PRD} estimated for LISA), the requirement on the TTL factors can be relaxed by five times to $1.0\times 10^{-4}$ m/rad ($= 0.10$ pm/nrad).

\begin{figure}[htbp]
    \centering
    \includegraphics[width=0.45\textwidth]{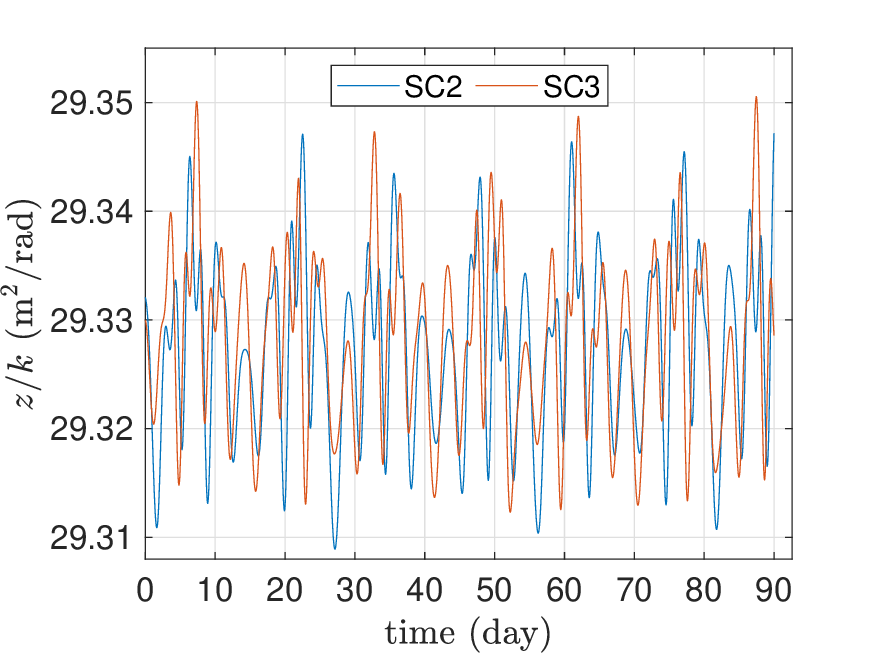}
    \caption{The time variation of the coefficient $z/k$ calculated from the remote satellite traces in the beam frame (cf. Fig. \ref{fig:trace}).}
    \label{fig:G}
\end{figure}

In Eqs. (\ref{eq:TTLalpha}, \ref{eq:TTLbeta}), the TTL factors are expressed as the products of the coefficient $z/k$ and the partial derivatives $\partial w_e/\partial x$ and $\partial w_e/\partial y$. From the remote satellite traces shown in Fig. \ref{fig:trace}, the coefficient $z/k$ can be calculated as a function of time (see Fig. \ref{fig:G}). The maximum value of $z/k$ is found to be less than 29.36 m$^2$/rad, and it implies that the partial derivatives of the wavefront deviation $|\partial w_e/\partial x|$ and $|\partial w_e/\partial y|$ at far field should be maintained below $3.4\times 10^{-6}$ rad/m.

To help confirm the far-field wavefront requirement derived from Eqs. (\ref{eq:TTLalpha}, \ref{eq:TTLbeta}), we take the defocused Gaussian beam as an example. Defocus is one of the main contributors to far-field wavefront distortion through diffraction. From \cite{Robertson97,LISA98}, in the case of defocus only, the corresponding phase noise can be expressed as
\begin{equation} \label{eq:defocusWFE}
    \delta \phi = \frac{1}{32} \left(\frac{2\pi}{\lambda}\right)^3 d D^2 \theta_0 \delta \theta, 
\end{equation}
where $d$ denotes the wavefront error at the exit pupil, and $D$ denotes the telescope diameter (30 cm for TianQin). Moreover, the total pointing error $\theta$ can be decomposed into the bias $\theta_0$ (DC and out-of-band slow-varying components) and the in-band fluctuation $\delta\theta$.

\begin{figure}[htbp]
     \centering
     \includegraphics[width=0.45\textwidth]{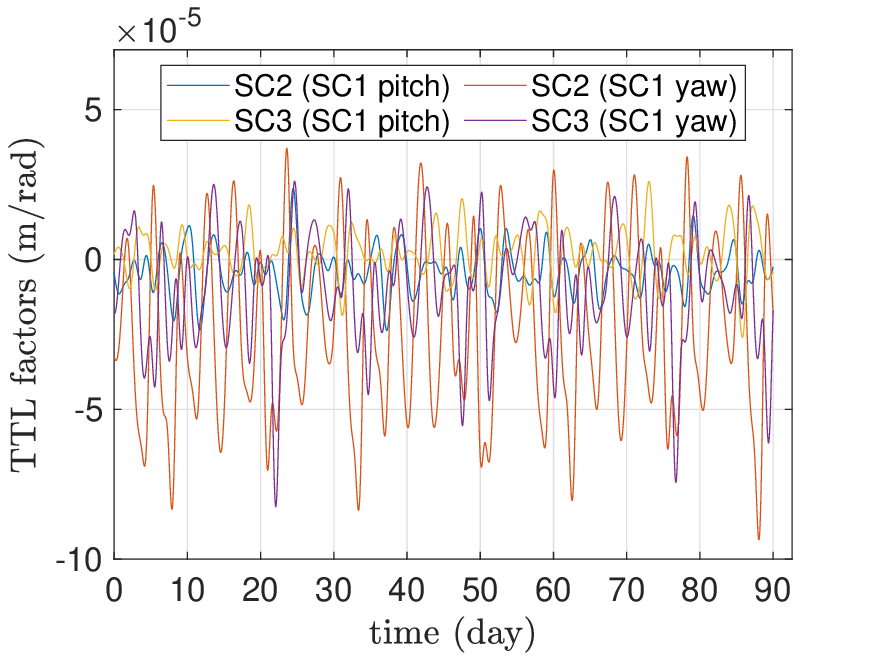}
     \caption{TTL factors with respect to the pitch and yaw angles ($\delta\alpha$, $\delta\beta$) for the arms SC1-2 and SC1-3, and with a $\lambda/37$ defocus and a 10 nrad pointing bias. }
     \label{fig:TTLfacxy}
\end{figure}

In the case of TianQin, when considering the uncompensated residual PAA (25 nrad) and the DC pointing bias (10 nrad), $\theta_0$ can attain a maximum value of 35 nrad. To fulfill TianQin's allocated budget (1 pm/Hz$^{1/2}$ before post-processing), the acceptable local wavefront error of the transmitted beam is estimated to be $d \approx \lambda / 37$ according to Eq. (\ref{eq:defocusWFE}). Adding this aberration to the transmitted Gaussian beam, we can obtain an analytical expression of $w_e$ based on the method of \cite{winkler1997}: 
\begin{equation}
w_{e}\! \left(x,y,z\right)\approx\frac{k \pi^{2} d \mathrm{D}^{2} \sqrt{\frac{z^{2}}{x^{2}+y^{2}+z^{2}}}}{8 \lambda^{2}}.
\label{eq:ff_wfe}
\end{equation}
Hence we estimate the maximum partial derivatives $\left| \partial w_e / \partial x\right| \approx \left| \partial w_e / \partial y\right| \approx 3.37 \times 10^{-6}\mathrm{rad}/\mathrm{m} $. This is consistent with the requirement of $3.4\times 10^{-6}$ rad/m derived from Eqs. (\ref{eq:TTLalpha}, \ref{eq:TTLbeta}) in previous paragraphs. Additionally, Fig. \ref{fig:TTLfacxy} shows the corresponding TTL factors throughout one 3-month observation window using the satellite traces of Fig. \ref{fig:trace}, which fall below  $1.0\times 10^{-4}$ m/rad requirement. Unfortunately, the analysis for general cases with other types of aberrations becomes much more complicated. This is deferred to future works.

\section{Concluding Remarks} \label{sec:conclusion}

In the paper, we have proposed for TianQin a fixed offset compensation strategy for PAAs based on the characteristics of TianQin's geocentric orbits. Moreover, we discuss related effect of the far-field wavefront errors for TianQin, and a corresponding requirement of $<3.4\times 10^{-6}$ rad/m is derived on the spatial derivatives of the far-field phase deviation $w_e$ from ideal Gaussian beams (see Eq. (\ref{eq:phase})). The analysis gives a preliminary green light for the proposed strategy. Regarding the future implementation, two points deserve to be mentioned.

First, since the TianQin satellites will be installed with single-side flat-top sun shields which must be illuminated by the Sun \cite{Zhang2018,Chen2021,Wang2023}, it requires that the three satellites perform a flip maneuver in attitude after each 3-month observation window. The maneuver interchanges the positions of the two MOSAs with respect to the normal of the constellation plane, hence flipping the parities of the PAA offsets as well. Therefore, the optical bench and telescope design should take into account these two scenarios, and be capable of switching between the two PAA pointing states.

Second, TianQin has proposed a 3+3 month observation scheme in its operation \cite{Luo2016}. To reduce the time spent on re-establishing laser links before each observation window or from accidental link losses, one option is to have a steering mirror performing the dual roles of fast link acquisition (see, e.g., \cite{HECHENBLAIKNER2023}) and PAA compensation. This means that once the scanning and acquisition phase is complete, the actuated mirror should switch to static PAA compensation. Then the process involves one transferring the pointing authority of the transmitted laser beam from the steering mirror to the DFAC and MOSA tracking control when the DWS signals become available, i.e., from in-field pointing to telescope pointing. During this process, the mirror can be controlled to gradually settle to a fixed angle for PAA offset in the science mode.

Since the pointing of the outgoing beam is static relative to the telescope and optical bench, barring small residual jitters from the mirror mechanism, some implications of the proposed PAA control scheme are the following. First, it may help ease the telescope design regarding the far-field wavefront error due to very small fields of view ($\pm 35$ nrad) required for the transmitted beams in the science mode. Second, it may render the total TTL factors less varying, and hence alleviating potential complexity in TTL estimation, and reducing possible data quality degradation due to PAA adjustments. More details are deferred to future works.

\begin{acknowledgments}
The authors thank Yuzhou Fang, Qing Xiao, Fan Zhu, Bobing Ye, Hsien-Chi Yeh, and Jun Luo for helpful discussions and comments. X. Z. is supported by the National Key R\&D Program of China (Grant Nos. 2022YFC2204600 and 2020YFC2201202) and NSFC (Grant No. 12373116), and Fundamental Research Funds for the Central Universities, Sun Yat-sen University (Grant No. 23lgcxqt001). 
\end{acknowledgments}

\bibliography{apssamp}

\end{document}